# Compensation of dipolar-exciton spin splitting in magnetic field

A. V. Gorbunov and V. B. Timofeev

*Institute of Solid State Physics RAS, Chernogolovka, Moscow Region, 142432, Russian Federation*
E-mail: gorbunov@issp.ac.ru

**Abstract.** Magnetoluminescence of spatially indirect dipolar excitons collected in 25 nm GaAs/AlGaAs single quantum well within a lateral potential trap has been studied in Faraday geometry. In the range of low magnetic fields, $B \leq 1.5$ T, an unusually large quadratic blue shift (upto 2 meV/T$^2$) is observed, which is related to in-plane electric field component $F_\parallel$ present in nonhomogeneous electrostatic field $F$ of the trap. The paramagnetic spin splitting of the luminescence line of the heavy-hole excitons in the trap centre is completely compensated at magnetic field below critical value $B_c \approx 2$ T. In the field range $B > B_c$ linear doublet splitting is observed for circularly polarized spin components of the opposite sign. The effect of spin-splitting compensation is caused by the exchange interaction in dense exciton Bose gas which is in qualitative agreement with the existing theoretical concepts.

## 1.    Introduction

Spin degrees of freedom in excitonic Bose systems and the exchange interaction between the spin-oriented components of the exciton ensemble become apparent in polarization of emitted irradiation. Analysis of the latter enables to considerably enhance understanding of the properties of the system in question. Specifically, in the case of microcavity exciton-polariton Bose condensate the phenomenon of Zeeman spin-splitting suppression in magnetic field below some critical value, $B < B_c$, has been recently predicted theoretically [1] and later confirmed experimentally [2, 3]. The effect is related to the spinor nature of exciton polaritons in microcavities: optically active heavy-hole excitons possess two spin projections onto the growth axis of the structure: $S_z = \pm 1$. Accordingly, luminescence light propagating along the axis contains two components circularly polarized in opposite directions: $\sigma^+$- and $\sigma^-$. In the absence of both magnetic field and interparticle interaction the components are energetically degenerate. When taking into account the exchange interaction the state of exciton Bose condensate becomes energetically favorable with equal numbers of bosons with up and down spins. It means that the luminescence light of exciton Bose condensate must be linearly polarized. If the system has no preferential direction, spontaneous symmetry breaking may take place which is manifested in a random-time varying direction of linear polarization. But in most experiments the objects under study are characterized by structural anisotropy which results in pinning of the linear polarization to a definite crystallographic direction. In particular, in GaAs/AlGaAs quantum wells grown in the (001) plane two mutually transverse directions, [110] and [1$\underline{1}$0], have a priority. Linear polarization along <110> has been observed both in exciton-polariton Bose condensate [2, 4] and in condensate of spatially indirect, dipolar excitons [5]. Two orthogonal linearly-polarized components are split in energy (usually by a small value of $\sim 10^{-5}-10^{-4}$ eV) and the linear polarization results from the primary occupation in the Bose condensate of the lowest-energy state. At zero magnetic field, with increasing optical pumping, the linear polarization degree $\rho_{lin}$ grows stepwise near the threshold of Bose-Einstein condensation (BEC), $\rho_{lin}$ decreasing gradually with further rise of pumping above the threshold because of condensate depletion due to the heating of the system. Magnetic field switching results in appearance of circularly-polarized $\sigma^+$- and $\sigma^-$-components, each of them being connected with a state where the spins are co-parallel and aligned with the magnetic field or opposed to it. According to theory [1], at thermodynamic equilibrium and zero temperature Zeeman energy splitting of the components is suppressed until the state with anti-parallel spins remains energetically more favorable than the state with co-parallel spins: the polariton-polariton exchange interaction exactly compensates Zeeman splitting, whereas the luminescence light remains elliptically polarized. With increasing magnetic field $B$ a critical value $B_c$ is achieved, when the energy of one of the spin-oriented states in the magnetic field becomes lower than the energy of the state with anti-parallel spins: there appears ($\sigma^+$–$\sigma^-$)-splitting proportional to the increment of magnetic field $\Delta B = B - B_c$. Indeed, suppression of Zeeman splitting at small magnetic fields has been observed experimentally in photoluminescence of exciton-polariton condensate [2, 3] in spite of the obvious non-equilibrium. Quite recently, compensation of spin-splitting in magnetic field for circularly polarized components due to the spin-anisotropic polariton-polariton interaction has been observed in a rectangular-pillar microcavity characterized by a large initial splitting of the linearly-polarized components (~200 μeV) and by a relatively weak deviation from equilibrium [6]. In this case the compensation effect was found to enhance with increasing polariton condensate density.

Unlike the exciton polariton in a microcavity the exciton itself in GaAs/AlGaAs heterostructures has a fourfold spin degeneracy. Besides an optically active exciton with spin $S_z = \pm 1$, there exists an optically non-active dark exciton with $S_z = \pm 2$. For the latter optical transitions are forbidden in the dipole approximation. The properties of the "four-component" exciton Bose condensate in magnetic field have been recently analyzed theoretically [7]. In particular, predicted was the feasibility of phase transitions in magnetic field between condensate states composed of different numbers of components. It means that the polarization behavior of condensed excitons in magnetic field may noticeably differ from that of cavity exciton-polariton condensate.

In this work we have experimentally studied the effect of perpendicular magnetic field $B_\perp$ on spatially indirect dipolar excitons in a 25-nm GaAs/AlGaAs quantum well collected in an electrostatic potential trap nearby a round window in the top Schottky gate (overview of the experimental results on exciton Bose condensation in such a trap can be found in [8, 9]). The spatially nonhomogeneous electric field of the trap contains not only perpendicular component $F_\perp$, normal to the surface, but also lateral one $F_\parallel$, parallel to the plane of the quantum well. It is well known that in crossed magnetic $\boldsymbol{B}$ and electric $\boldsymbol{F}$ fields excitons move along direction $\boldsymbol{F} \times \boldsymbol{B}$, i.e. the dispersion curve for excitons $E(\boldsymbol{k})$ shifts in the momentum space so that excitons with finite momentum $\boldsymbol{k} \neq 0$ possess minimal energy. As a result, optical transition with photon emission becomes indirect in the momentum space and one should expect an increase of the exciton radiative lifetime and, hence, deeper exciton cooling. For spatially-indirect interwell excitons subjected to homogeneous perpendicular electric $F_\perp$ and parallel, in-plane, magnetic field $B_\parallel$ the effect has been studied both theoretically [10, 11] and experimentally [12-14]. The theory of direct intrawell exciton in homogeneous parallel electric $F_\parallel$ and perpendicular magnetic $B_\perp$ fields has been developed in [15, 16].

In the present work indirect excitons are experimentally explored in crossed fields, homogeneous perpendicular magnetic field $B_\perp$ and inhomogeneous parallel electric field $F_\parallel$. Owing to the symmetry, in the vicinity of the round window in the Schottky gate the in-plane component of static electric field $F_\parallel$ has only radial constituent $F_r$, which is maximal near the window edge and decays down to null in the window centre. It is worth noting that the geometry of the crossed fields with radially symmetric in-plane electric field $F_r$ and a normal to the plane magnetic field $B_\perp$, causes the excitons to move along the ring trajectories around the window centre. In other words, a magnetoelectric trap for spatially-indirect excitons can be realized. Essentially the same idea, with the use of a point contact on top of a cylindrical sample instead of a window in the metal film, has been previously proposed in [17] as a fairly promising approach to experimental implementation of exciton Bose condensation.

## 2. Experimental technique

Spatially-indirect dipolar excitons were studied in a single wide (25 nm) GaAs quantum well at electric field normal to heterolayers, applied between the metal Schottky gate on top of the AlGaAs/GaAs-heterostructure and the conducting electron layer inside the structure (built-in electrode) [18]. Due to the applied electric field the dipolar excitons exhibited a large dipole moment (over 100 D). In the system under study dipole-dipole repulsion prevented the excitons from binding into exciton molecules and other multi-particle complexes. Both photoexcitation and photoluminescence detection was done through a round window ø7 μm in the opaque metal layer of the Schottky gate (100 nm thick Au/Cr film). Dipolar excitons were collected in a circular lateral potential trap that appeared along the window perimeter owing to the strongly nonhomogeneous electric field [18, 19]. The sample inside the superconducting magnet was placed in luquid $^4$He in an optical cryostat that enabled experiments in the range of magnetic fields $0 < B < 6$ T at $T \approx 1.6$ K. Luminescence light was collected by means of a fused-silica lens (focal length $F = 12.5$ mm, numerical aperture $NA \approx 0.4$) positioned inside the cryostat.

Dipolar excitons were simultaneously excited with two continuous wave lasers with wavelengths $\lambda_{sb}$ =782 nm (photoexcitation with photon energy below the energy gap in the AlGaAs barrier, "subbarrier" excitation) and $\lambda_{ob} = 659$ nm ("overbarrier" excitation). The laser spot diameter on the sample surface was ~30 μm. Combining the two laser photoexcitations allowed to achieve maximum possible compensation of extra charges inside the trap and to maintain the exciton system as close to neutrality as possible [18, 20]. The architecture of the used structures and lateral traps as well as the details of charge compensation are discussed in [8, 9, 18].

The magnified image of the round window in the Schottky gate, used both for excitation and detection of photoluminescence, was projected on the entrance slit of the spectrometer (focal length 500 mm) equipped with a cooled silicon CCD-camera. The use of an imaging spectrometer transferring the image from the plane of the entrance slit to that of the exit slit without aberrations enabled: a) to record the luminescence spectra with spatial resolution along the direction of the spectral slit, and b) to register, in the zeroth order of diffraction grating, images of the sample in different spectral ranges with the help of a narrow band (≈0.7 nm) interference filter. The optical projective system used to transfer the sample image from the cryostat allowed observation of real-space patterns of dipolar exciton photoluminescence with spatial resolution of ≤2 μm.

All the measurements of the luminescence spectra were carried out with spatial resolution. A narrow central stripe from the image of the window in the Schottky gate passing through the entrance slit of the spectrometer and dispersed into the spectrum by diffraction grating, was registered as an image on the sensitive matrix of the CCD camera. Luminescence spectra from the top/bottom edge of the window and its centre were obtained. The luminescence polarization was analyzed by means of a Glan prism and a quarter-wave phase plate.

## 3. Experimental results and discussion

The circularly polarized photoluminescence spectra versus magnetic field both at the edge of the window in the Schottky gate and in its centre are presented in Fig.1. The right ($\sigma^+$) and left ($\sigma^-$) circularly polarized components are shown as solid and dashed lines, respectively. At the chosen value of the electric field, $F_\perp \approx 7.6$ kV/cm, the width of the spectral line of dipolar excitons without magnetic field (FWHM) amounts to $\Delta E \geq 0.8$ meV both at the edge and in the centre of the window. With switched magnetic field the lines narrow noticeably: at B $\geq 1.5$ T $\Delta E$ equals to $\approx 0.5$ meV at the edge and $\approx 0.35$ meV in the centre. The intensity of the exciton luminescence line in the centre of the window is stronger than at the edge (compare the left and right vertical scales in Fig.1), while the Stark shift to the low-energy side is less by 7.5 meV since the electric field strength $F_\perp^*$ is lower in this region. Besides the strong heavy-hole-exciton (hh) line, a much weaker (30÷100-fold) line of light-hole exciton (lh) as well as lines of the higher-energy excited states of both hh- and lh-excitons are observed at the window centre.

Fig.2 shows the behavior of exciton-line energy $E$ as a function of magnetic field. In the range of low fields, $B_\perp \leq 1.3$ T, an unusually large quadratic-in-field blue shift of the exciton lines is observed: $\approx 2.2$ meV/T$^2$ for exciton at the window edge, $\approx 1.9$ meV/T$^2$ for hh-exciton and $\approx 0.9$ meV/T$^2$ for lh-exciton at the window centre. These values exceed, by at least an order of magnitude, the normal Langevin diamagnetic shift, which is determined in this case by the in-well area of the dipolar exciton and stays typically below 0.1 meV/T$^2$ in GaAs structures [21]. The "giant" diamagnetic shift is not observed in the flat-band regime, at $F_\perp \approx 0$. It is also absent in homogeneous perpendicular electric field [20]. Therefore, it is reasonable to connect the effect with the presence of the radial field component in the plane of the quantum well in the nonhomogeneous electric field near the window. It can be assumed that in crossed magnetic and electric fields the central symmetry of the lowest exciton state gets broken, higher states with larger angular moments being admixed. The magnitude of radial component $F_r$ depends on radial coordinate $r$. In pure electrostatics $F_r(r)$ is maximal in the vicinity of the window edge due to the effect of the electric field concentration at the metal edge. When moving away from the edge, the component decays monotonously down to zero exactly in the window centre: $F_r(0) = 0$. However, the actual distribution of $F_r(r)$ is difficult to calculate: one should take into account field screening by photogenerated charge carriers. At the window edge $F_r$ can be expected to be of the same order of magnitude as $F_\perp$, and it should decrease considerably when moving to the centre. Since the value of the applied external electric field is sufficiently large ($\approx 7.6$ kV/cm), the presence of the lateral component should be also appreciable inside the window.

At $B_\perp \geq 1.3$ T dependence $E(B)$ is close to linear (see Fig.2a). The slope of the straight line for hh-exciton both at the edge and in the centre of the window amounts to $\approx 0.81$ meV/T, while for lh-exciton in the centre it is slightly less. Obviously, in the limit of strong magnetic field energy $E(B)$ is determined by the behavior of the lowest Landau level for exciton: $E_o = \hbar eB/2\mu c = \hbar\omega_c/2$, where reduced mass $\mu = m_e m_h/(m_e + m_h)$, $m_e$ and $m_h$ are the electron and hole masses, respectively, and $\omega_c = eB/\mu c$ the cyclotron frequency. But here, in the range of moderate magnetic field, exciton binding energy $E_b$ and cyclotron energy $\hbar\omega_c$ are of the same order. That is why the value of the effective reduced mass extracted from the slope of dependence $E(B)$, $\mu_{eff} \approx 0.071 m_o$, is definitely overestimated: it exceeds the electron mass in GaAs $m_e \approx 0.066 m_o$, (here $m_o$ is the free electron mass).

Spin- (paramagnetic) splitting in the magnetic field of the line of indirect hh-exciton into $\sigma^+$- and $\sigma^-$- components occurs both in the centre and at the edge of the window, but in quite a different way. Fig.2 shows the correspondent dependencies of Zeeman splitting $\Delta E_Z = E_{\sigma^-} - E_{\sigma^+}$ as functions of magnetic field $B_\perp$.

At the window edge the splitting grows linearly in the range of magnetic field $0 \leq B \leq 1.5$ T, i.e. $\Delta E_Z = \mu_B g B$, where $\mu_B$ is the Bohr magneton, and the effective g-factor of an exciton is $g_x \approx +0.9$ (the energy of the $\sigma^-$-component is higher). With $1.5 \leq B \leq 4$ T the splitting value remains more or less constant, $\Delta E_Z \approx 0.07$ meV, while with further field increase it diminishes and approaches zero. This spin-splitting behavior is fairly common in wide GaAs quantum wells. It has been previously observed in the experiments both with a non-doped GaAs/AlGaAs quantum well 25 nm wide in the absence of electric field [22] and in a weakly doped 30 nm GaAs/AlGaAs quantum well in a perpendicular electric field [23]. The dependence with a small spin-splitting value and a change of the g-factor sign in moderate magnetic fields has been described theoretically in the effective-mass approximation, with regard for valence band mixing of hole states [24].

On the contrary, Zeeman splitting for hh-excitons in the middle of the window is quite unusual: it appears to be compensated (with an accuracy of ±20 µeV) at low magnetic fields, below $B \approx 2$ T. At $B > 2$ T $\Delta E_Z$ grows linearly with field, while $g_x \approx -1.5$ (in this case the energy of the $\sigma^+$-component is higher). Simultaneously, in the same place of the trap, the spin splitting for the lh-excitons increases monotonically with $B \geq 0$ (see Fig.2c). It is characterized by an effective g-factor, $g_x \approx +7$, just as it occurs in a 25-nm AlGaAs/GaAs quantum well under homogeneous photoexcitation without any lateral trap [20].

Thus, in spite of the fact that Bose-condensation effects have been previously observed exactly for hh-excitons collected in the ring electrostatic trap *at the edge of the window* in the Schottky gate [8, 9], no indication of compensated Zeeman splitting for excitons in magnetic field was found. It should be noted that spectrally selective registration of the luminescence images shows a considerable change in the spatial distribution of the dipolar-exciton luminescence at the window edge when the magnetic field is turned on (see Fig.3). In the absence of magnetic field (Fig.3a) a typical pattern of two pairs of bright spots located symmetrically in vertices of a square with diagonals along directions [110] and [1$\underline{1}$0] is observed (see [8, 9, 18]) (in our case, due to insufficient spatial resolution, the pattern is close to a square with sides parallel to [100] и [$\underline{1}$00]). In magnetic field (Fig.3b) the spots spread along the perimeter of the window and a square with sides along [110] and [1$\underline{1}$0] is observed. Therefore, this suggests that the exciton motion along the circular trajectory in crossed perpendicular magnetic field $B_\perp$ and radial electric field $F_r$ destroys the conditions for exciton accumulation in an annular lateral trap up to the density sufficient for BEC.

In the middle of the window, on the contrary, without magnetic field no sign of exciton Bose condensation has been previously found [18]. With magnetic field, however, just in the window centre exciton spin splitting behaves exactly as it was predicted for spinor Bose condensate [1]: there is no splitting in low magnetic field, $0 \leq B \leq B_c \approx 2$ T, while at $B > B_c$ it grows linearly with field increment $\Delta B = B - B_c$. The observed 2.5-fold narrowing in the magnetic field of the hh-exciton luminescence line points to the possibility of attaining a high exciton density that is sufficient to form degenerate gas of interacting Bose particles. Note that the external electric field used in the present work is almost by an order of magnitude stronger than in the previous studies [8, 9, 18]. Hence, the effective field strength $F_\perp$ in the middle of the window is also notably higher. So, the excitons in this region are also spatially indirect though with a smaller dipole moment than at the edge. Besides, just in the vicinity of the window centre a magnetoelectric trap, similar to that that proposed in [17], in which dipolar excitons are collected by means of winding in crossed fields, might be realized. Clearly, at the same conditions the concentration of lh-excitons is lower by 1-2 orders of magnitude. As a result in this, much less dense, exciton system no degeneracy is achieved, the collective effects are weak and no unexpected spin slitting is observed.

Circular polarization degree, $\rho_{circ} = (I_{\sigma+} - I_{\sigma-})/(I_{\sigma+} + I_{\sigma-})$, where $I_{\sigma+}$ and $I_{\sigma-}$ are the intensities in the line maximum at $\sigma^+$- and $\sigma^-$-polarization, respectively, behaves in magnetic field quite similarly for dipolar hh-excitons in the middle and at the edge of the window (see Fig.2d): $\rho_{circ}$ increases monotonously with magnetic field from 0 up to ≈0.2, in both cases the $\sigma^+$-component being more intense. But there is also a remarkable difference: in the case of spin splitting the stronger $\sigma^+$-component always has a higher energy than the $\sigma^-$-component in the middle of the window, and always a lower energy at the edge. There is obviously no thermodynamic equilibrium in the exciton spin system at the window centre, the degree of nonequilibrium increasing with magnetic field. At the window edge the more intense $\sigma^+$- component has a lower energy than the σ⁻-component, i.e. the system is closer to equilibrium. It is not surprising because the lifetime of the indirect exciton at the window edge was found to exceed that at the centre by at least an order of magnitude [25]. Nevertheless, in sufficiently strong magnetic fields the spin equilibrium in this region is also destroyed, as $\rho_{circ}$ increases with magnetic field while the value of splitting $\Delta E_Z$ decreases. In the same conditions the light-hole exciton line behaves, at least qualitatively, as it should be at equilibrium: the low-energy σ⁺-component is more intense and polarization degree $\rho_{circ}$ grows with increasing Zeeman splitting $\Delta E_Z$.

In zero magnetic field the luminescence line of indirect excitons in a circular electrostatic trap at the window edge is linearly polarized along direction <110> in the plane of quantum well {001}. As shown previously [5], linear polarization degree $\rho_{lin}$ is maximal at the condensation threshold (>70%) and, with further increase of optical pumping, diminishes gradually due to condensate depletion caused by system heating. In our case, with the chosen experimental parameters in the absence of a magnetic field polarization degree, $\rho_{lin} = (I_{110} - I_{1\underline{1}0})/(I_{110} + I_{1\underline{1}0})$, where $I_{110}$ and $I_{1\underline{1}0}$ are the intensities in the line maximum with polarization along [110] and [1$\underline{1}$0], respectively, at the window edge $\rho_{lin} \approx 0.2$, the luminescence in the window middle being unpolarized, $\rho_{lin} \approx 0$. In the explored range of magnetic field, $0 \leq B \leq 6$ T, the linear polarization degree of dipolar exciton photoluminescence remains constant both at the edge and in the centre of the window.

## 4. Conclusions

Upon accumulation of spatially-indirect dipolar excitons in a wide GaAs/AlGaAs quantum well inside an electrostatic trap formed by the nonhomogeneous electric field nearby the window in the Schottky gate suppression of spin splitting has been found in a perpendicular to quantum well magnetic field (Faraday geometry) in the field range below $B_c \approx 2$ T. The effect is related to Zeeman splitting compensation due to the exchange interaction in dense degenerate gas of Bose particles with nonzero spin. Besides, at low magnetic fields, $B \leq 1.3$ T, an anomalously large quadratic-in-field high-energy shift is observed, up to $\approx 2.2$ meV/T², which is connected with the presence of a considerable lateral component of the (parallel) electric field in the quantum well plane. In crossed fields, perpendicular magnetic and axially symmetric radial electric fields, exciton motion along the circular trajectories around the window centre is possible which may result in mixing of the centrally-symmetric lowest-energy state of the exciton with higher states possessing larger angular moments.

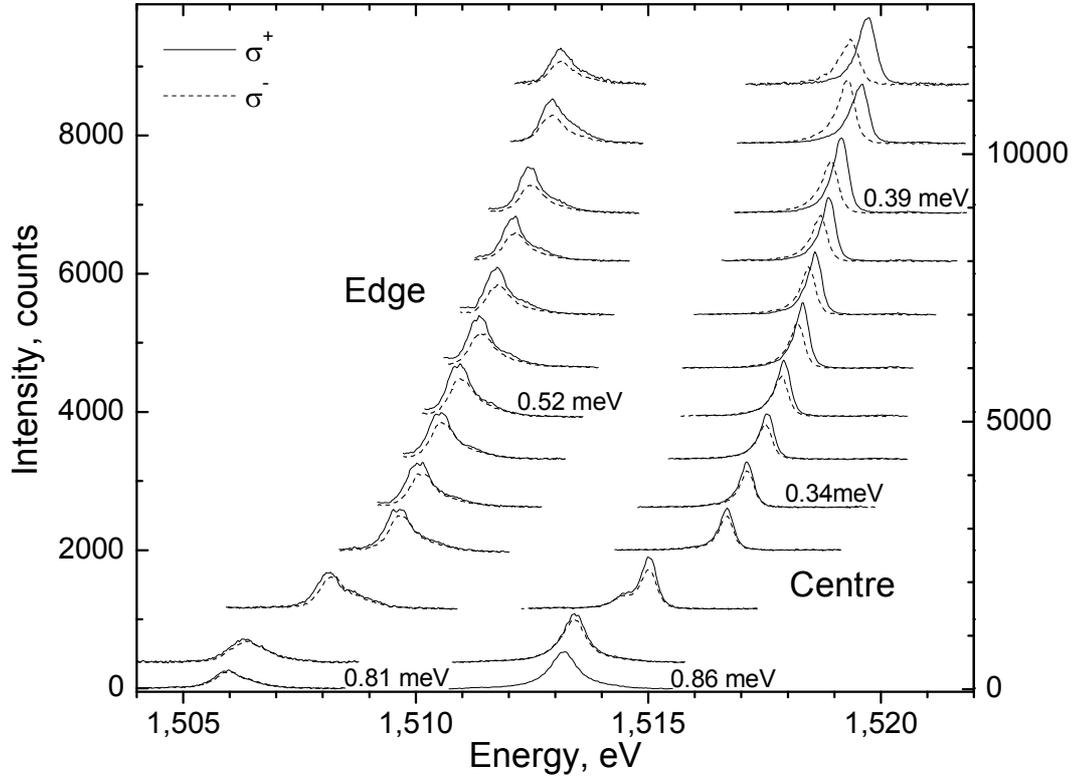

Fig.1. Circularly polarized photoluminescence spectra at the window edge (left series) and in the middle of the window (right series) versus magnetic field $B$ increasing from 0 to 6 T with step 0.5 T from bottom to top. Some linewidths (FWHM) are shown. Photoexcitation power is $P_{ob}$ = 11.5 μW for overbarrier laser and $P_{sb}$ = 118 μW for sub-barrier laser. Temperature $T$ = 1.6 K.

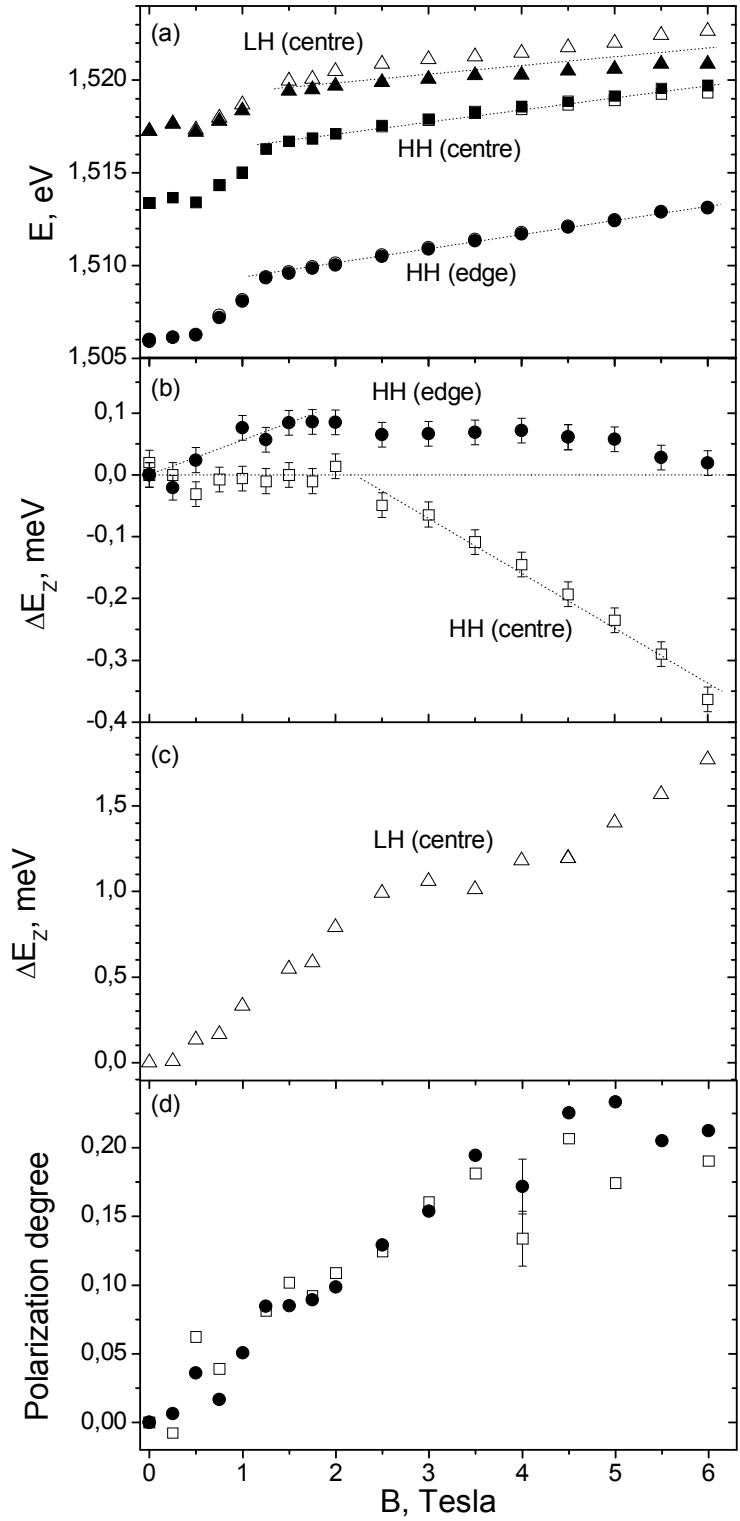

Fig.2. Energy (a), Zeeman splitting (b,c) and circular-polarization degree (d) vs magnetic field $B$ for the luminescence line of dipolar heavy-hole exciton at the edge (circles) and in the middle of the window (squares), as well as for the light-hole exciton in the middle (triangles). Solid symbols in Fig.2a correspond to $\sigma^-$- and open ones to $\sigma^+$- circularly polarized components, respectively.

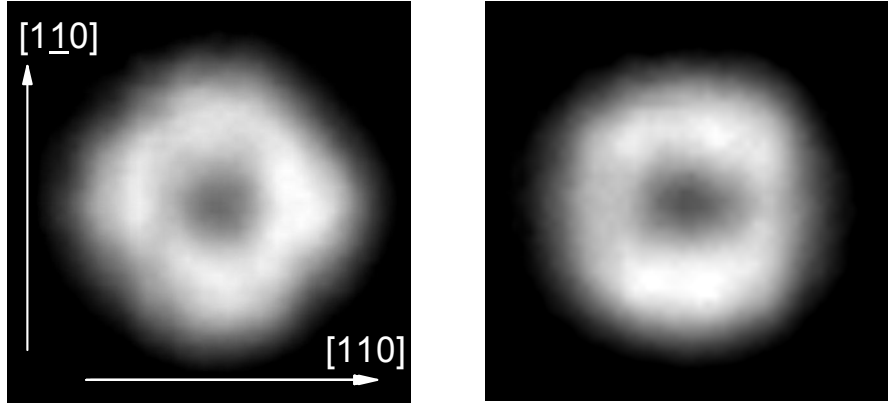

Fig.3. Spatial distribution of dipolar-exciton luminescence in the circular electrostatic trap at the edge of the window in the Schottky gate: (a) in the absence of magnetic field, $B_\perp = 0$, and (b) in magnetic field $B_\perp = 3$ T.